\documentclass[12pt]{iopart}

\newcommand{\be}{\begin{eqnarray}}
\newcommand{\ee}{\end{eqnarray}}
\newcommand{\bea}{\begin{eqnarray}}
\newcommand{\eea}{\end{eqnarray}}

\usepackage{graphicx}
\usepackage{bm}
\usepackage{color}
\usepackage{slashed}
\usepackage{cite}
\usepackage{graphicx}
\usepackage{multicol}
\usepackage{graphicx}
\usepackage{color}
\usepackage{slashed}
\usepackage{CJKutf8}
\begin{document}
\begin{CJK}{UTF8}{<font>}
\title{Rastall gravity: accretion disk image in radiation fields context and visual transformations compared to Reissner-Nordstr\"{o}m black holes}

\author{Yu-Xiang Huang$^{1}$, Sen Guo$^{*2}$, Yu Liang$^{3}$, Yu-Hao Cui$^{3}$, Qing-Quan Jiang$^{*1}$, Kai Lin$^{*3}$}

\address{$^1$School of Physics and Astronomy, China West Normal University, Nanchong 637000, People's Republic of China}
\address{$^2$Guangxi Key Laboratory for Relativistic Astrophysics, School of Physical Science and Technology, Guangxi University, Nanning 530004, People's Republic of China}
\address{$^3$Hubei Subsurface Multi-scale Imaging Key Laboratory, Institute of Geophysics and Geomatics, China University of Geosciences, Wuhan 430074, People's Republic of China}

\ead{$^1$ yxhuangphys@126.com \\ $^2$ sguophys@126.com  \\ $^3$ washy2718@outlook.com \\ $^3$ yuhaocui@cug.edu.cn \\ $^1$ qqjiangphys@yeah.net  \\ $^3$ lk314159@hotmail.com }
\vspace{10pt}
\begin{indented}
\item[]Dec 2023
\end{indented}

\begin{abstract}
Our study investigates  the astronomical implications of Rastall gravity, particularly its behavior amidst a radiation field compared to Reissner-Nordstr"{o}m (RN) black holes. Our research delineates a crucial correlation between the dynamics of the accretion disk and the parameters $Q$ and $N_{\rm r}$, which aptly reflect the influence of spacetime metrics on the disk's appearance. Elevated electric charge $Q$ prompts contraction in the disk¡¯s orbit due to enhanced gravitational effects, while higher $N_{\rm r}$ values lead to outward expansion, influenced by the radiation field's attributes. Interestingly, the charged black holes surrounded by radiation fields display distinct visual disparities from RN black holes. Brightness decreases and expansion occurs within the accretion disk's innermost stable circular orbit with rising $N_{\rm r}$ values. Our study also reveals the process by which the accretion disk transitions from a conventional disk-like structure to a hat-like form at different observation angles, with the redshift effect gradually intensifying. Moreover, the results of the Rastall gravity radiation field we consider are consistent with the constraints of the host galaxy's gravitational lensing on the Rastall gravity parameters, enhancing the consistency between theoretical predictions and actual observations.

\end{abstract}


\section{Introduction}
\label{intro}
\par
The covariant conservation law of the energy-momentum tensor plays a paramount role within the Einstein's framework of general relativity (GR). P. Rastall introduced a theory that posits potential modifications to GR, incorporating a proportionality constant denoted as $\kappa$. This constant establishes a linkage between the Einstein tensor and the energy-momentum tensor, leading to a transformation of the Einstein field equation to $G_{\rm \mu\nu}+\kappa {\rm \lambda} g_{\mu\nu}R =\kappa T_{\rm \mu\nu}$ \cite{1}. Recently, scholarly attention has increasingly focused on exploring Rastall's theory. Notably, static and spherically symmetric solutions within Rastall gravity have been employed to describe neutron stars, imposing more stringent constraints on deviations from GR as posited by Rastall \cite{2}. Rastall's theory has revealed a new category of black hole solutions, encompassing both uncharged and charged Kiselev-like black holes \cite{3}. Moreover, Moradpour and colleagues initiated an investigation into traversable wormholes conforming to Rastall's theory. Their inquiry involved exact solutions with baryonic matter states, revealing the theory's pronounced impact on wormhole characteristics, intricately tied to the Rastall dimensionless parameter and initial conditions \cite{4}. Additionally, Kumar and co-researchers extended upon the Kerr-Newman black hole solution within the context of Rastall gravity. This extension facilitated an analysis of thermodynamic attributes of rotating Rastall black holes, investigating the interplay between particle motion and effective potentials \cite{5}. These studies collectively underscore the growing intrigue surrounding the ramifications of Rastall gravity. However, a research gap persists regarding the observable attributes of astrophysical systems as dictated by Rastall theory. Addressing this void, our investigate the discernible properties of an accretion disk encircling a black hole enshrouded by a perfect fluid within the ambit of Rastall gravity.

\par
The potential displacement of Einstein's theory by Rastall theory has sparked an enduring confluence of contention and discourse within the domain of gravitational physics. Investigations led by Visser have proposed a perspective wherein, in a majority of instances, Rastall gravity can be interpreted as a meticulous rearrangement of the matter facet within the confines of Einstein gravity's framework, thereby implying a certain interchangeability of these two gravitational constructs \cite{6}. Visser's analysis thus sheds light on the potential analogies and correlations that underlie these two theories. By orchestrating this reconfiguration of the matter domain, Rastall gravity introduces an alternate lens through which to perceive the intrinsic nature of gravitational interplays. This divergence of viewpoints poses a challenge to the conventional comprehension of Einstein's theory and its foundational presumptions. On the contrary, Darabi and collaborators posit an alternative narrative, asserting that Rastall gravity represents an ``open'' paradigm distinct from the tenets of Einstein gravity \cite{7}. This perspective emphasizes a clear demarcation between these two theoretical frameworks. Their perspective accentuates the distinctive attributes inherent to Rastall gravity in contrast to Einstein gravity. According to their portrayal, Rastall gravity introduces novel constituents and characteristics that set it apart from the well-established edifice of Einstein's theory. Due to the ``open'' character of Rastall gravity, a prospect emerges for the emergence of novel physical phenomena, thereby beckoning further exploration and scrutiny. Consequently, considerable scholarly endeavor has been channeled into delineating the potential equivalence of these two theories \cite{8,9,10,11,12}.

\par
The existence of accretion disks surrounding black holes is a well-substantiated phenomenon supported by a wealth of empirical evidence. A significant milestone was reached with the accomplishments of the Event Horizon Telescope (EHT), which provided unparalleled insights into black hole structure. Notably, the iconic depiction of a supermassive black hole within the Messier 87$^{*}$ (M87$^{*}$) galaxy captured a photon ring of luminosity surrounding the black hole, enveloped by an accretion disk \cite{13}. This phenomenon is mirrored by the supermassive black hole Sagittarius A$^{*}$ (Sgr A$^{*}$) situated at the core of the Milky Way Galaxy, where the existence of an accretion disk in the vicinity of the black hole has been observed \cite{14}. The inception of accretion disk models traces back to the 1970s, notably exemplified by the Shakura-Sunyaev model portraying a geometrically thin, optically thick disk that characterizes its optical attributes and thickness \cite{15}. Expanding on this foundation, Luminet employed a semi-analytical approach to explore the primary and secondary images of an accretion disk around a Schwarzschild black hole \cite{16}. Laor contributed a formulation for calculating line profiles emitted from accretion disks around rotating black holes, highlighting the preponderance of flux emanating from the disk's inner realm and manifesting a blue shift in the line peak \cite{17}. The depiction of accretion disks in various modified gravity contexts of black holes has evoked extensive inquiry, extensively elucidated in the scholarly discourse \cite{18,19,20,21,nnn}. Delving into the Rastall theory's framework for understanding black holes enriches comprehension of this alternative gravitational paradigm. Guo and colleagues, in particular, have undertaken a systematic exploration of the photon ring and slim accretion disk properties in charged black holes within Rastall gravity \cite{22,23}. Their findings underscore the intricate relationship between accretion morphology, accretion disk positioning, and the optical manifestations of black holes.

\par
Galaxies are widely acknowledged to comprise two distinct constituents: stellar components and dark matter components. The stellar segment finds an effective description through a perfect fluid model, a characterization driven by its minimal stickiness, negligible shear stresses, and perceptible visibility \cite{eee}. In a corresponding manner, a model positing dark matter as a perfect fluid has emerged in recent years, notably introduced by Rahaman and colleagues. This model portrays dark matter as a non-sticky, non-shear stressed, and inherently non-visible perfect fluid \cite{fff}. Consequently, in the investigation of galaxies within the context of Rastall's gravity, the comprehensive adoption of a perfect fluid representation for the entirety galactic matter emerges as a prudent methodology. This approach not only aligns with the inherent attributes of the constituent components but also serves as a pivotal strategy in delineating and constricting the gravitational parameters pertinent to Rastall's theory through astronomical observations.

\par
The exploration of accretion disk properties around black holes serves as a crucial avenue for deepening our comprehension of the fundamental physical mechanisms governing these intricate systems. This study is dedicated to advancing our comprehension through an investigation into the observable attributes of thin accretion disks within the paradigm of Rastall gravity. The aim of this research is to elucidate the distinctive impact on the observable properties of accretion structures that arises from modifications to the geometric characteristics of spacetime within the Rastall gravity paradigm. This paper is organized as follows. In Section \ref{sec:2}, our investigation focused on the accretion disk of a black hole enveloped by a perfect fluid within the framework of the Rastall theory. The trajectory of the black hole accretion disk enveloped by a radiation fluid was successfully simulated using the numerical integration method. This allowed us to obtain both the primary and secondary images of the observed radiant flux. Furthermore, we conducted a fitting analysis to establish the relationship between the parameters of the black hole and the characteristics of the most stable circular orbit and the photon ring. In Section \ref{sec:3}, we constrained the parameters in Rastall¡¯s gravitational theory by utilizing observational data from 118 strong gravitational lens systems in galaxies, thereby determining the range of values for $\kappa \lambda$. Conclusions and discussions are presented in Section \ref{sec:4}.

\section{Trajectory and radiation intensity}
\label{sec:2}
\subsection{Ray trajectory}
\label{sec:2-1}
\par
The metric of perfect fluid envelopment in the Rastall theory is given by \cite{3}
\begin{equation}
\label{2-1}
{\rm d}s^{2}=-f(r){\rm d}t^{2}+\frac{1}{f(r)}{\rm d}r^{2}+r^{2}{\rm d}\theta^{2}+r^{2}\sin^{2}\theta {\rm d}\phi^{2},
\end{equation}
where $f(r)$ represents the metric of a black hole, and its specific form is:
\begin{equation}
\label{2-2}
f(r)=1-\frac{2M}{r}+\frac{Q^{2}}{r^{2}}-\frac{N_{\rm r}}{r^{\xi}},
\end{equation}
where $M$ is black hole mass, $Q$ is black hole charge, and $N_{\rm r}$ is the radiation field parameter. The parameter $\xi \equiv \frac{1+3\omega-6\kappa\lambda(1+\omega)}{1-3\kappa\lambda(1+\omega)}$ in which $\lambda$ is the Rastall parameter, $\kappa$ is the Rastall gravitational coupling constant, $\omega$ is the equation of state parameter.

\par
In our examination, we utilize the Novikov-Thorne model for thin accretion disks to analyze the presence of such a disk encircling the black hole. A comprehensive explanation of this model is available in the citation \cite{16}. For an observer situated at a distance, we compute the cumulative alteration of the ray bending angle, leading to the subsequent outcomes, i.e.
\begin{equation}
\label{2-3}
\psi (u) = \int_{u_{\rm source}}^{u_{\rm obs}}\frac{{\rm d}u}{\Omega(u)}=\int_{u_{\rm source}}^{u_{\rm obs}}\frac{{\rm d}u}{\sqrt{\frac{1}{b^{2}}-u^{2}f(u)}},
\end{equation}
where the parameter $u$ defined as $u \equiv \frac{1}{r}$. The integration limits are associated with the photon's emission point denoted as $u_{\rm source}$ and the observer's position denoted as $u_{\rm obs}$. In cases where specific ray impact parameters surpass a critical threshold, a radial inflection point denoted as $u_{0}$ could be evident. Under such circumstances, photon movement can be segregated into two distinct segments: from $u_{0}$ to $u_{\rm source}$ and from $u_{\rm obs}$ to $u_{0}$. The value of the radial inflection point is determined as $u_{0}=\frac{1}{b^{2}f(r)}$ within these mentioned segments. To replicate the primary and secondary images of the black hole, numerical integration techniques are employed. The trajectories of photons are calculated using the subsequent integration formula:
\begin{equation}
\label{2-4}
\int_{u_{\rm source}}^{u_{\rm obs}}\frac{{\rm d}u}{\sqrt{\frac{1}{b^{2}}-u^{2}f(u)}}=n \pi-\arccos{\frac{\sin{\eta}\tan{\theta_{0}}}{\sqrt{\sin^{2}\eta\tan^{2}\theta_{0}+1}}}.
\end{equation}
Here, we denote the impact parameter as $b$, the celestial angle as $\eta$, and the observer tilt angle as $\theta_{0}$. The parameter $n$ corresponds to the image order, with $n=0$ and $n=1$ representing the direct and secondary images, respectively. In this equation, the observer's inclination can be directly determined, allowing the photon trajectory to depend on the impact parameter $b$, the distance $r$ of the orbit from the black hole, and the azimuthal angle $\eta$. When examining a specific orbit, $r$ can be directly set, thus converting the equation into a relationship between $b$ and $\eta$, which also includes parameters from the metric  (considered as constants). This equation enables a clear observation of how changes in parameters can lead to alterations in the spacetime geometry, ultimately affecting the appearance of the accretion disk. Direct integration to solve for $u$ poses certain challenges; however, by setting $\eta$ within the range of 0 to $2 \pi$, a root-finding approach can be employed to determine the corresponding $b$, thereby yielding the observed orbital imagery.

\par
This study focuses on the optical characteristics of an accretion disk around a black hole, using a specific perfect fluid radiation field mentioned in reference \cite{3,aaa,bbb} as an illustrative example. The radiation field under consideration in this investigation is characterized by a specific equation of state parameter, denoted as $\omega=\frac{1}{3}$. The transformation of Eq. (\ref{2-2}) yields the following expression, we have
\begin{equation}
\label{2-5}
f_{r}(r)=1-\frac{2M}{r}+\frac{Q^{2}-N_{\rm r}}{r^{2}}.
\end{equation}
When $N_{\rm r}=0$, it returns to RN black hole. Hence, our study encompasses the scenario of RN black holes, which aids in distinguishing between Einstein gravity and Rastall gravity. Typically, when $f(r)=0$, two roots are present, namely the outer event horizon radius ($r_{+}$) and the inner event horizon radius ($r_{-}$). By non-dimensionalizing $Q$ and $N_{\rm r}$ and obtaining the relationship between $q=\frac{Q}{M}$ and $n=\frac{N_{r}}{M^{2}}$, the admissible ranges for $Q$ and $N_{\rm r}$ can be determined to ascertain the presence of a black hole \cite{23}.

\par
Figure 1 illustrates the direct and secondary images of the accretion disk at radii $r=9M$, $17M$, and $25M$, while taking into account a variety of parameter configurations. For $Q=0.5$, and varying $N_{\rm r}$ as $N_{\rm r}=0$ (RN black hole) and $N_{\rm r}=2$, a discernible outward expansion in the orbital path of the accretion disk is evident relative to its reference position. Conversely, when keeping $N_{\rm r}=1$ fixed and incrementing $Q$ to $Q=0.1$ and $Q=0.9$, an observable inward contraction of the accretion disk's orbit is observed. Due to variations in parameters leading to changes in spacetime geometry, thereby affecting the appearance and structure of the accretion disk, we conducted a highly intuitive comparative analysis of the direct orbital images at $r=9M$ to visually demonstrate this phenomenon. It is observed that an increase in $Q$ results in a stronger gravitational pull from the black hole, inducing a contraction of the accretion disk towards its center. Conversely, an increase in the value of $N_{\rm r}$ causes the disk to expand outward, attributed to the influence of the perfect fluid radiation field on the spacetime structure. This behavior highlights the crucial interplay between gravitational and radiation forces, determining the changes in spacetime geometry within the Rastall theory framework, leading to alterations in the appearance of the accretion disk. Furthermore, in the context of the Rastall theory, a detailed comparison between a charged black hole characterized by a perfect fluid radiation field and the established RN black hole reveals minimal observable distinctions. This observation suggests that, in this scenario, the impact of the Rastall theory on the visual features of the accretion disk, such as direct and secondary orbital images, remains relatively consistent with predictions made by the established RN black hole model.

\begin{center}
\includegraphics[width=5cm,height=5cm]{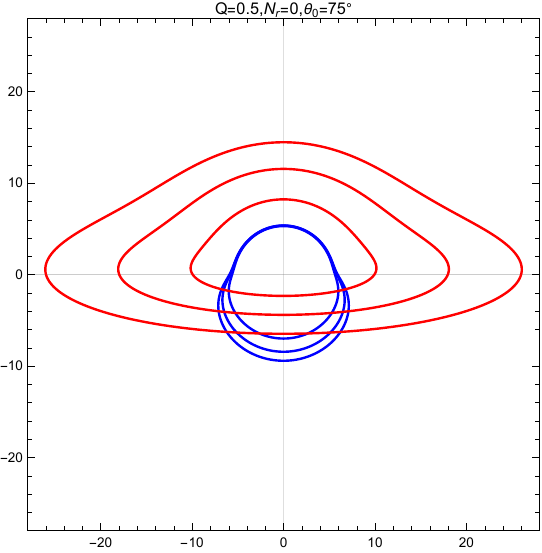}
\includegraphics[width=5cm,height=5cm]{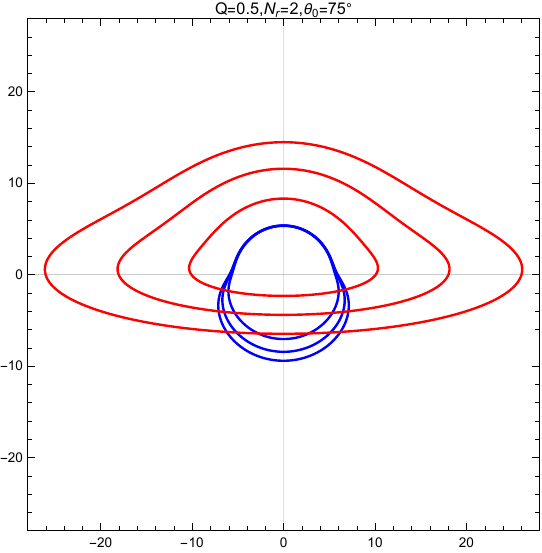}
\includegraphics[width=5cm,height=5cm]{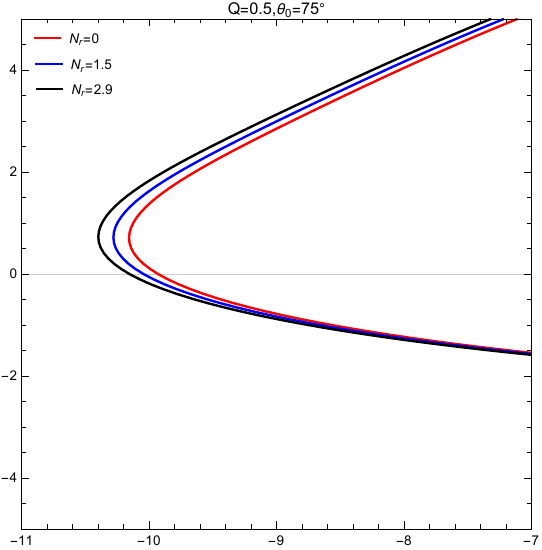}
\includegraphics[width=5cm,height=5cm]{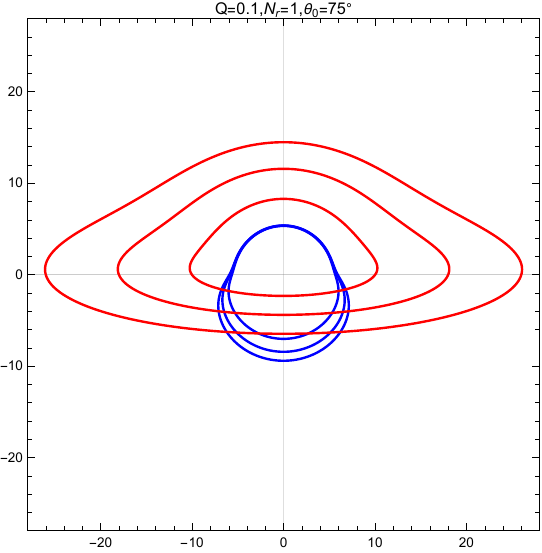}
\includegraphics[width=5cm,height=5cm]{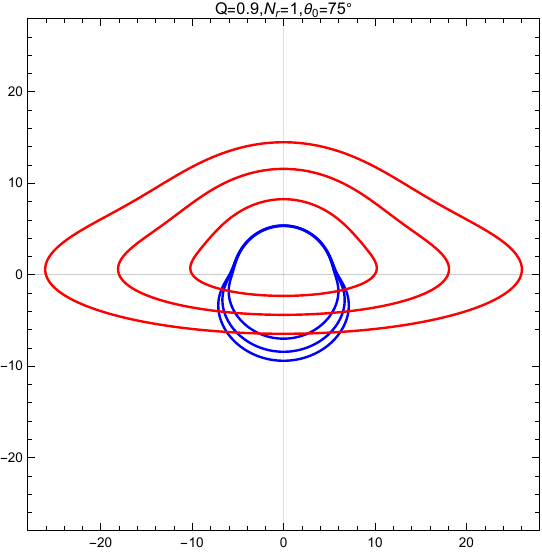}
\includegraphics[width=5cm,height=5cm]{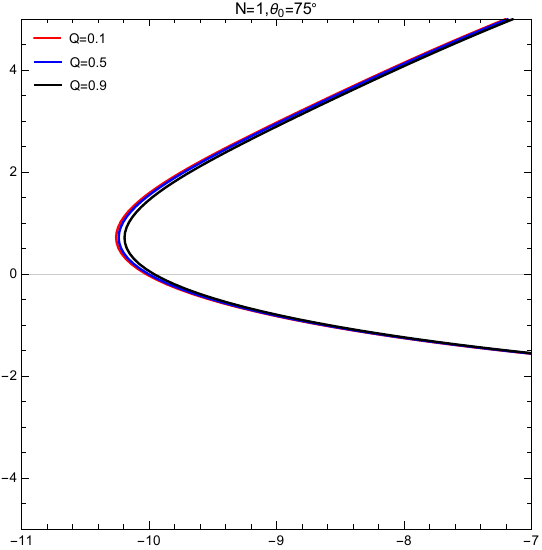}
\parbox[c]{15.0cm}{\footnotesize{\bf Fig~1.}
The direct (red line) and secondary (blue line) images of the black hole accretion disk at the observation angle ($\theta_{0}=75^{\circ}$). These lines represent distinct radial distances: $r=9M$, $r=17M$, and $r=25M$, with the innermost line corresponding to $r=9M$. In the top panel, the charge is fixed at $Q=0.5$, while concurrently varying the radiation field parameter $N_{\rm r}$ with values of $0$ and $2$ from left to right. The far right segment illustrates a comparative view of partial direct images for different $N_{\rm r}$ values at $r=9M$. In the bottom panel, $N_{\rm r}$ is fixed at 1, and the charge $Q$ is varied from left to right, taking values of $0.1$ and $0.9$. The far right segment showcases a comparative view of partial direct images for different $Q$ values at $r=9M$. The black hole mass is set to $M=1$.}
\label{fig1}
\end{center}

\subsection{Observable features of the thin disk}
\label{sec:2-2}
\par
The radiant flux of a thin accretion disk is examined in this subsection. The expression for the radiant flux is derived from previous studies \cite{24,25}.
\begin{equation}
\label{2-6}
F = - \frac{\dot{M}}{4\pi \sqrt{-g}} \frac{\Omega_{,\rm r}}{(E-\Omega L)^{2}} \int_{r_{\rm in}}^{r} (E- \Omega L)L_{,\rm r} {\rm d} r.
\end{equation}
In the equation, $\dot{M}$ denotes the mass accretion rate, $g$ represents the determinant of the metric, and $r_{\rm in}$ denotes the inner edge of the accretion disk. The variables $\Omega$, $E$, and $L$ correspond to angular momentum, energy, and angular velocity, respectively, and are expressed as
\begin{eqnarray}
\label{2-7-1}
&&E=-\frac{g_{\rm tt}}{\sqrt{-g_{\rm tt}-g_{\rm \phi\phi}\Omega^{2}}},\\
\label{2-7-2}
&&L=\frac{g_{\rm \phi\phi}\Omega}{\sqrt{-g_{\rm tt}-g_{\rm \phi\phi}\Omega^{2}}}, \\
\label{2-7-3}
&&\Omega=\frac{{\rm d}\phi}{{\rm d}t}=\sqrt{-\frac{g_{\rm tt,r}}{g_{\rm \phi\phi,r}}}.
\end{eqnarray}
Combining equations (\ref{2-2})-(\ref{2-7-3}), we obtain the radiation energy flux on the black hole.
\begin{equation}
\label{2-8}
F = \frac{\dot{M}(-8Q^{2}r^{\xi}+I)}{4\sqrt{2}\pi r^{4}X}\int_{r_{in}}^{r}\frac{r^{-3-\xi}(Y-N_{r}^{2}r^{4}\xi(2+\xi)-Z)}{\sqrt{2}X} {\rm d} r,
\end{equation}
where $X$, $Y$, $I$, and $Z$ are defined as $X=\sqrt{\frac{-2Q^{2}+2Mr+N_{\rm r}r^{\rm \xi-2}\xi}{r^{4}}}(2r^{\xi}(2Q^{2}+r^{2}-3Mr)-N_{\rm r}r^{2}(2+\xi))$, $Y=-2r^{2\xi}(4Q^{4}-9MQ^{2}r+M(6M-r)r^{2})$, $I=r(6Mr^{\rm \xi}+N_{\rm r}r\xi(2+\xi))$, and $Z=N_{\rm r}r^{2+\xi}(2Mr+Q^{2}(\xi-10)\xi-2Mr(\xi-6)\xi+r^{2}(\xi-2)\xi)$, respectively. In the context of a distant observer measuring the radiant flux, it becomes imperative to account for the influence of redshift. The observed flux can be accurately described by the following relationship \cite{16}.
\begin{equation}
\label{2-9}
F_{\rm obs} = \frac{F}{(1+z)^{4}}.
\end{equation}

\par
Within these parameters, the quantity $1+z$ corresponds to the redshift factor, an essential component that finds determination through a comparison between the energy of the photon emitted, denoted as $E_{\rm em}$, and the energy of the photon observed, denoted as $E_{\rm obs}$. The emitted energy $E_{\rm em}$ arises from an integral across the surface of the accretion disk, where both gravitational and Doppler shift effects are at play. In contrast, the observed energy $E_{\rm obs}$ emerges from consideration of the redshift attributed to the motion of the emitting source in relation to the observer. The connection linking the energies of emission and observation is aptly conveyed by $E_{\rm obs}=(1+z)E_{\rm em}$, where the redshift factor $z$ intimately hinges upon the gravitational potential and the velocity of the emitting source. This factor assumes utmost significance as it substantially informs the accurate assessment of observed flux and serves as a conduit to comprehend the characteristics of radiation discharged by the black hole system. The $E_{\rm em}$ is
\begin{equation}
\label{2-10}
E_{\rm em}=p_{\rm t} \mu^{\rm t} + p_{\rm \phi} \mu^{\rm \phi} = p_{\rm t} \mu^{\rm t} \Bigg(1 + \Omega \frac{p_{\rm \phi}}{p_{\rm t}}\Bigg),
\end{equation}
where $p_{\rm t}$ and $p_{\rm \phi}$ represent the photon 4-momentum. For observers located at a significant distance, the ratio $p_{\rm t}=p_{\rm \phi}$ signifies the impact parameter of the photons in relation to the z-axis. This relationship can be expressed in terms of trigonometric functions as follows:
\begin{equation}
\label{2-11}
\sin \theta_{0} \cos \alpha = \cos \gamma  \sin \beta,
\end{equation}
one can obtain
\begin{equation}
\label{2-12}
\frac{p_{\rm t}}{p_{\rm \phi}} = b \sin \theta_{0} \sin \alpha.
\end{equation}
Therefore, we can obtain the expression for the redshift factor
\begin{equation}
\label{2-13}
1 + z = \frac{E_{\rm em}}{E_{\rm obs}} = \frac{1 + b \Omega \cos \beta}{\sqrt{-g_{\rm tt} - 2 g_{\rm t \phi} - g_{\rm \phi\phi}}}.
\end{equation}

\par
By employing equations (8) and (9), images of a thin accretion disk surrounding a black hole, characterized by diverse parameters within the gravitational framework of Rastall, can be acquired. The primary focus of this inquiry is to attain a profound comprehension of black holes immersed in perfect fluid radiation fields, a paradigm offered as a demonstrative instance. As depicted in Figure 2, the observational representation is attainable in scenarios where a black hole is immersed in a perfect fluid radiation field typified by an equation of state parameter $\omega=\frac{1}{3}$. Evidently discernible from the figure is the trend that, with a constant parameter $Q$, an elevation in the value of $N_{\rm r}$ corresponds to a marked diminishment in the luminosity of the accretion disk. When $N_{\rm r}=0$, corresponding to the RN black hole, the observer observes the radiation flux reaching its maximum, and the innermost stable circular orbit is closest to the event horizon of the black hole. Simultaneously, the innermost stable circular orbit progressively extends as $N_{\rm r}$ increases. Conversely, when $N_{\rm r}$ is maintained at a constant value, an inverse pattern emerges, manifesting varied outcomes for the luminance of the accretion disk and the dimensions of the innermost stable circular orbit. These observations distinctly underscore the profound impact of the perfect fluid radiation field parameter $N_{\rm r}$ on the luminosity and structural stability attributes of the accretion disk that encircles the black hole.

\begin{center}
\includegraphics[width=5cm,height=4.2cm]{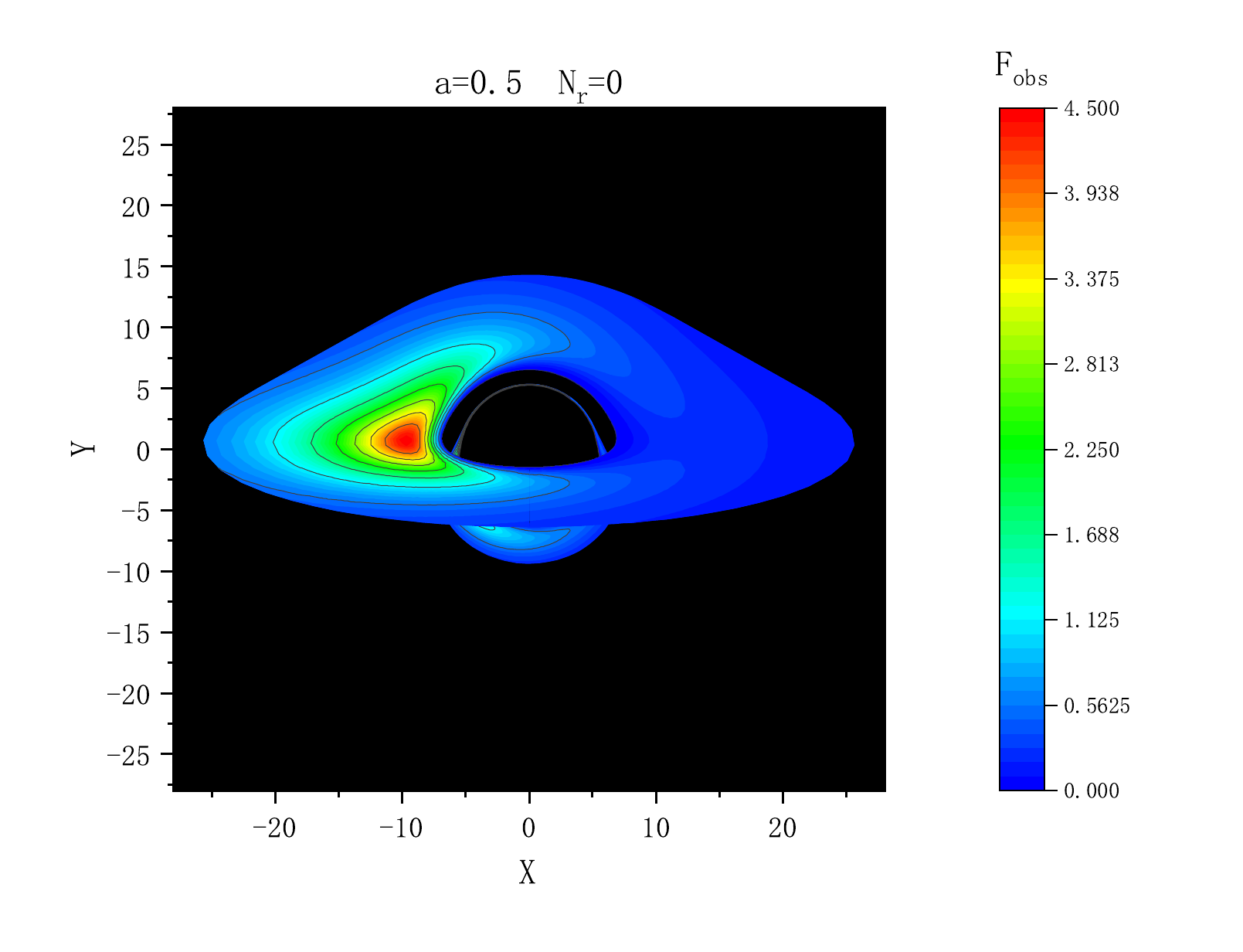}
\includegraphics[width=5cm,height=4.2cm]{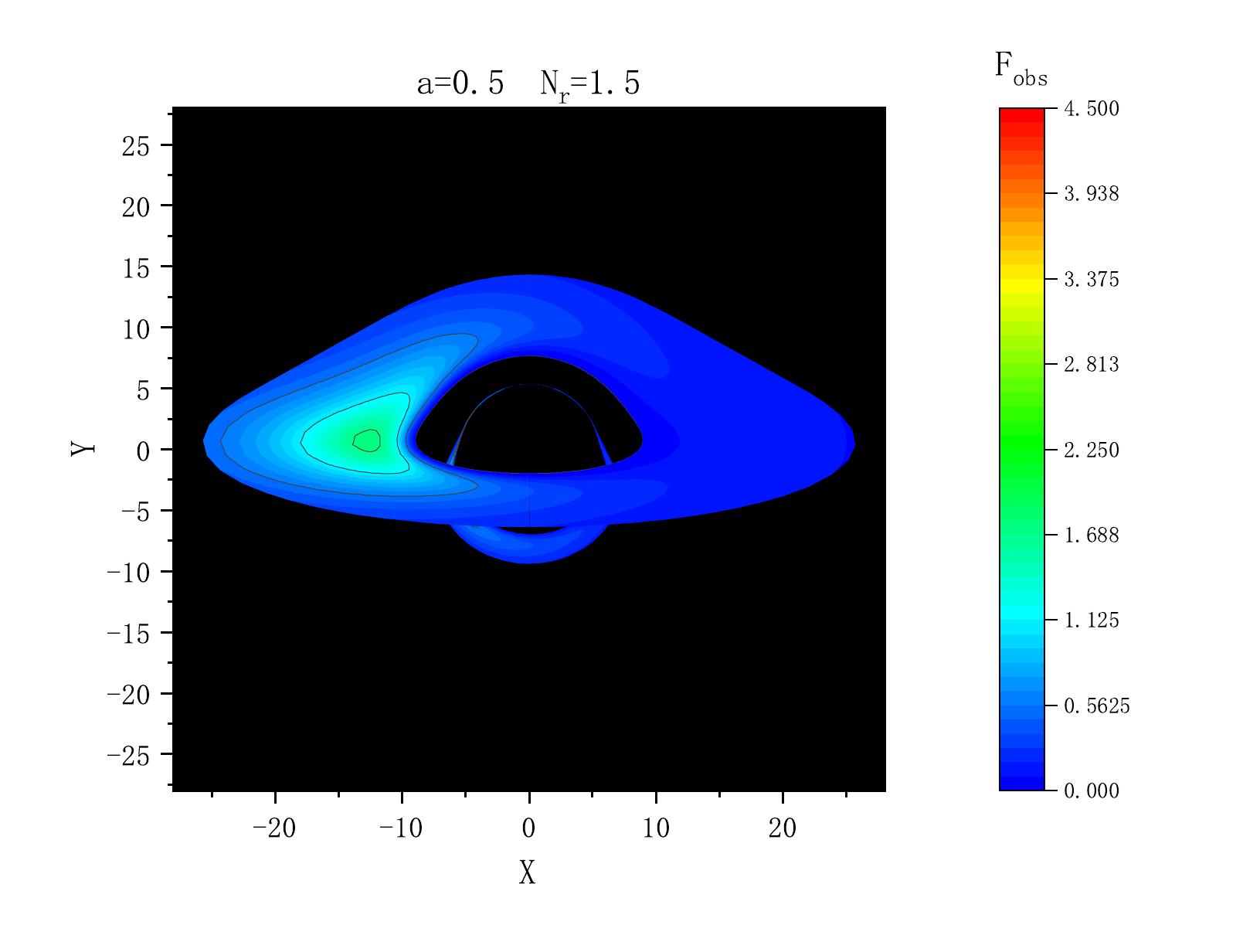}
\includegraphics[width=5cm,height=4.2cm]{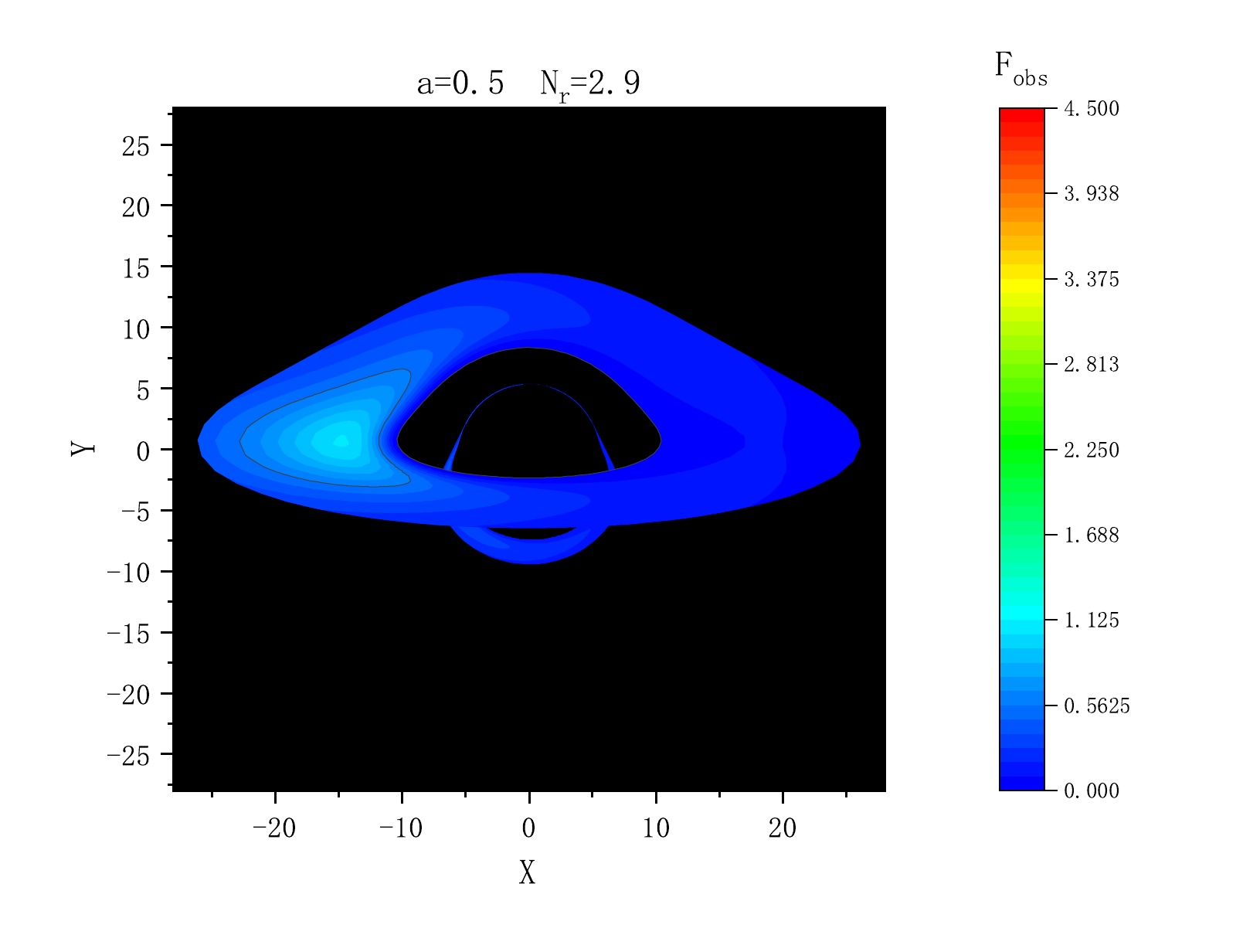}
\includegraphics[width=5cm,height=4.2cm]{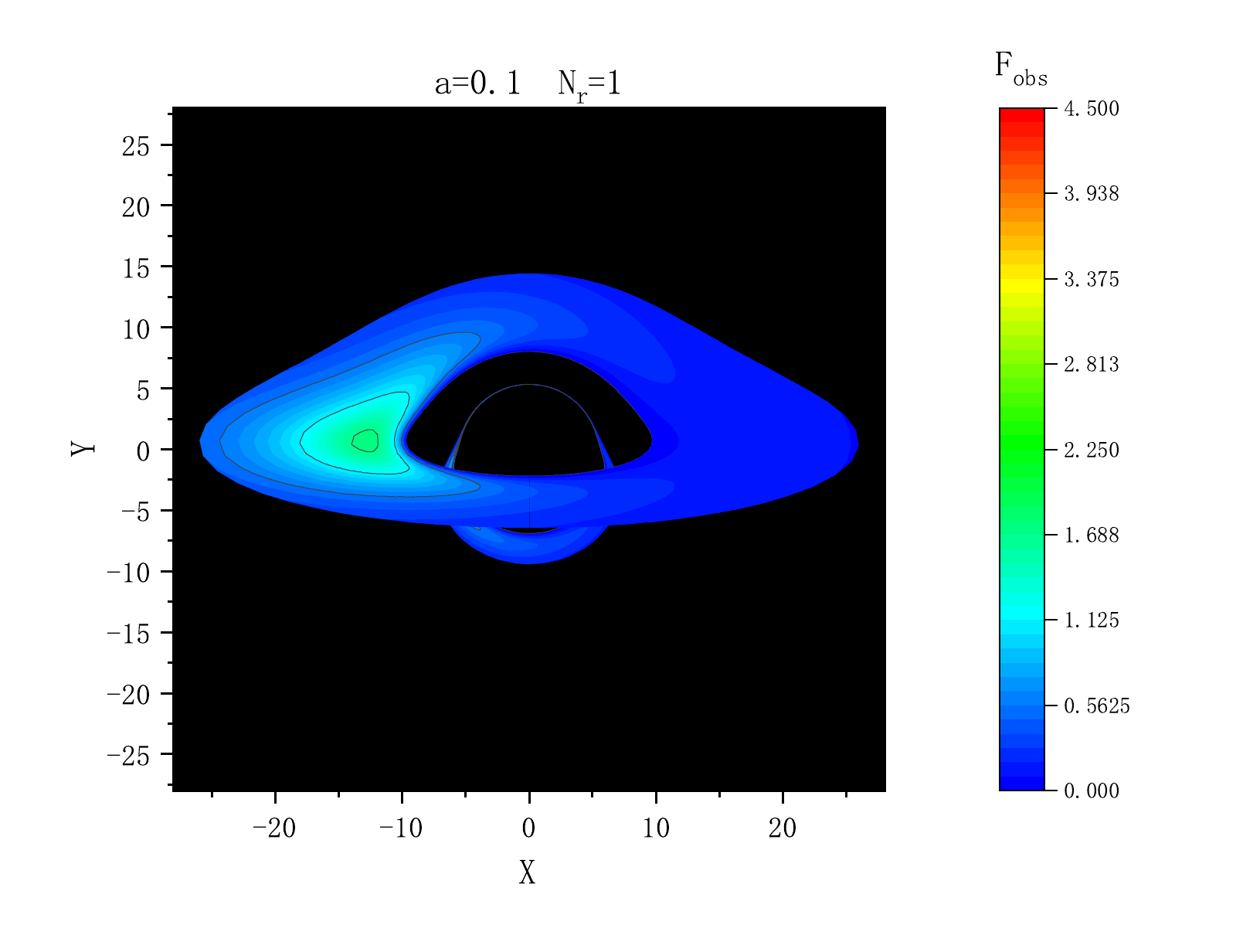}
\includegraphics[width=5cm,height=4.2cm]{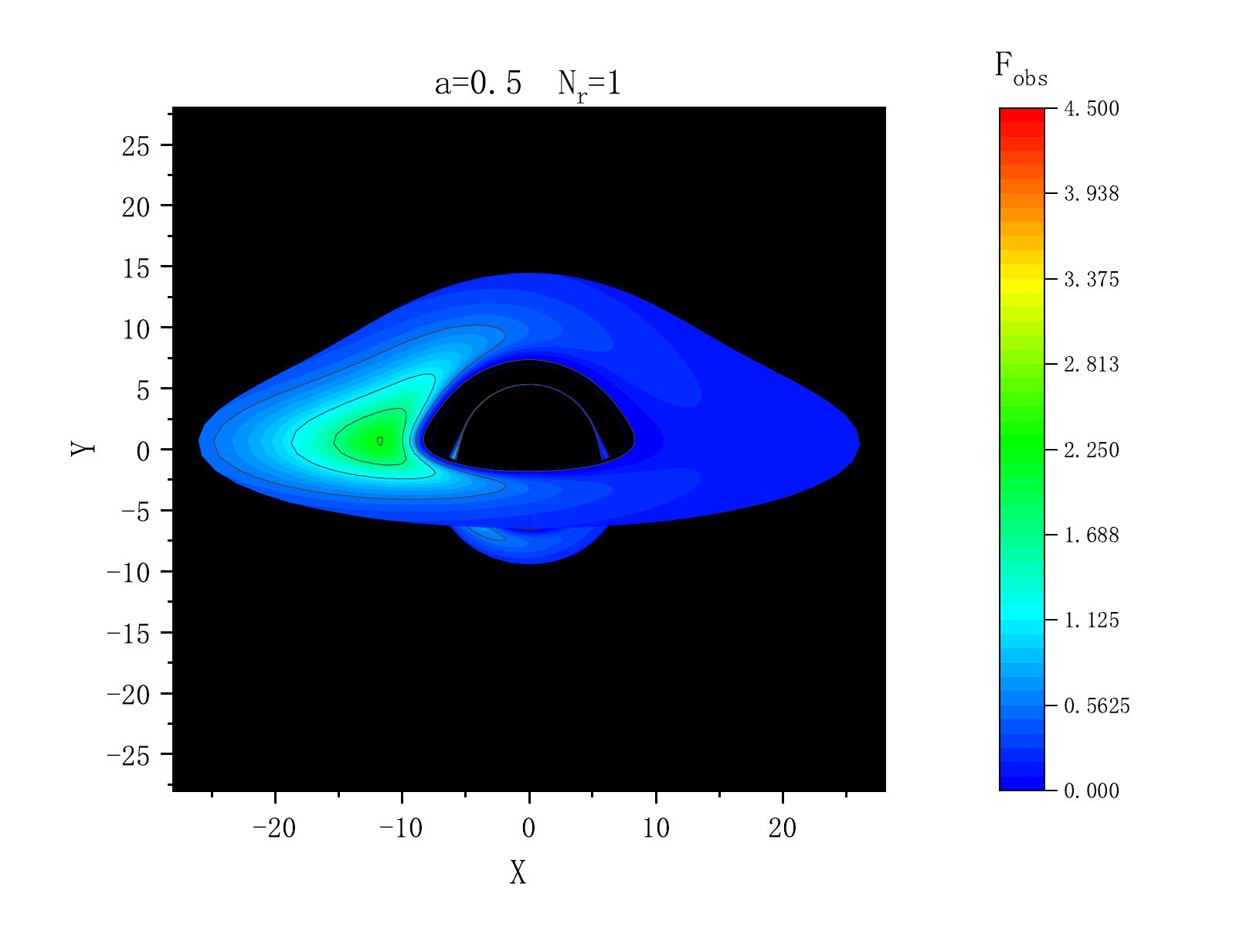}
\includegraphics[width=5cm,height=4.2cm]{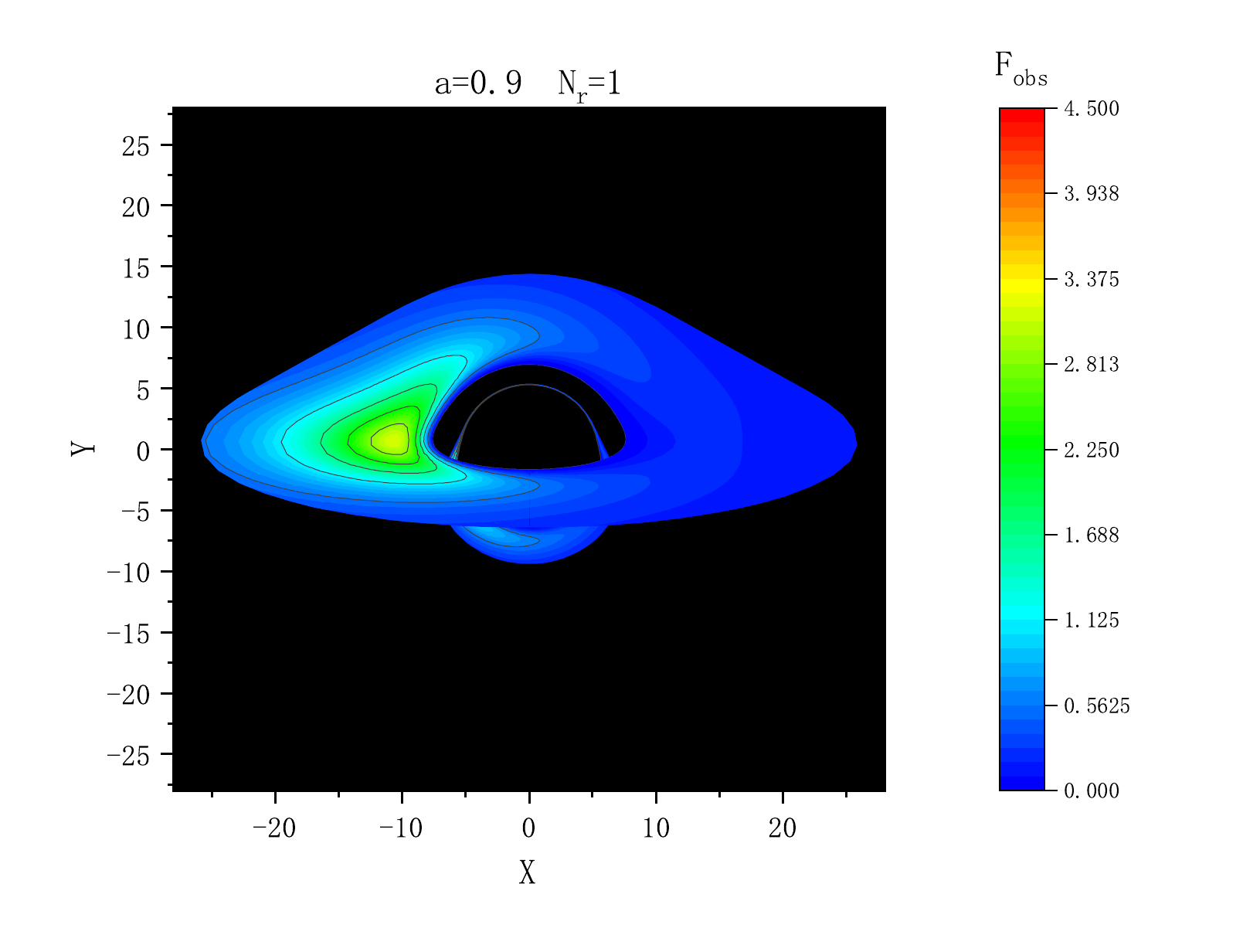}
\parbox[c]{15.0cm}{\footnotesize{\bf Fig~2.}  
The observed flux $F$ is presented at an inclination angle of $\theta_{0}=75^{\circ}$. The inner edge of the disk at $r_{\rm in} = r_{\rm isco}$, and the outer edge of the disk is at $r=25M$. In the top panel, we show the results for the charge $Q=0.5$ with varying radiation field parameters $N_{\rm r}$ from left to right: $0$, $1.5$, and $2.9$. In the bottom panel, we present the results for $N_{\rm r}=1$ with varying charge $Q$ from left to right: $0.1$, $0.5$, and $0.9$. The black hole mass is chosen as $M=1$.}
\label{fig2}
\end{center}

\par
Figure 3 illustrates the predicted observational images achieved by adjusting the observer's tilt angle while maintaining constants $Q=1$ and $N_{\rm r}=2.9$. As the observation angle progressively increases ($\theta_{0}=15^{\circ}$, $45^{\circ}$, and $75^{\circ}$), discernible alterations unfurl within the morphology of the accretion disk. A gradual metamorphosis transpires, wherein the disk-like structure gradually transitions into a configuration reminiscent of a hat. Moreover, the radiant flux intensity exhibits a progressively growing asymmetry between the left and right facets of the accretion disk. This emphasized asymmetry is rooted in the heightened influence of the Doppler effect as the observation angle increases. The origin of this effect can be traced to the relative motion between the observer and the matter emitted within the confines of the accretion disk. Consequently, this Doppler effect results in a more pronounced red or blue shift in the observed radiation, contingent upon the direction of motion. As a consequence, the disparity in radiant flux intensity between the left and right boundaries of the accretion disk magnifies appreciably with the augmentation of the observation angle. These predicted outcomes align with other research findings and may offer valuable guidance for achieving the scientific objectives of the EHT in future studies.

\begin{center}
\includegraphics[width=5cm,height=4.2cm]{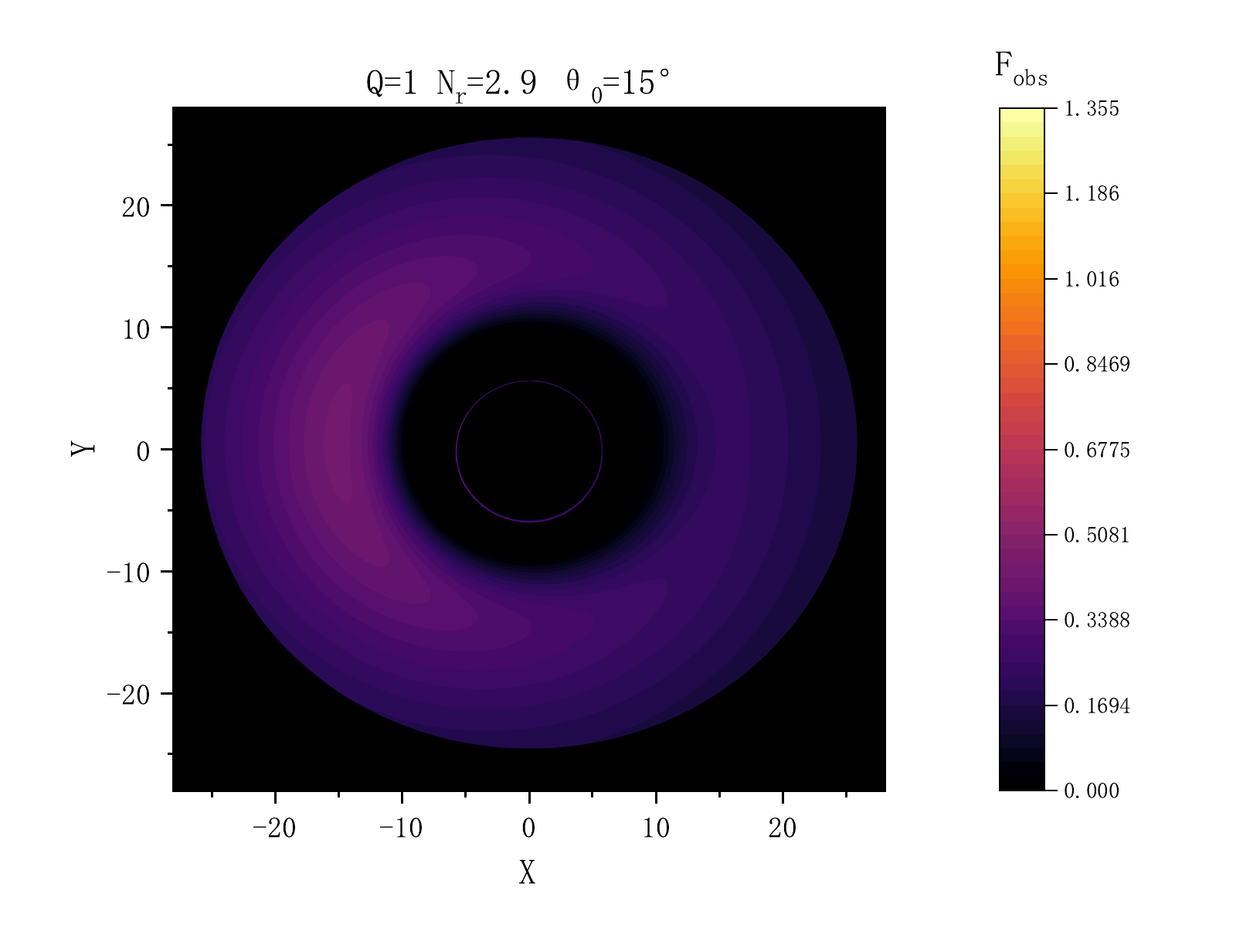}
\includegraphics[width=5cm,height=4.2cm]{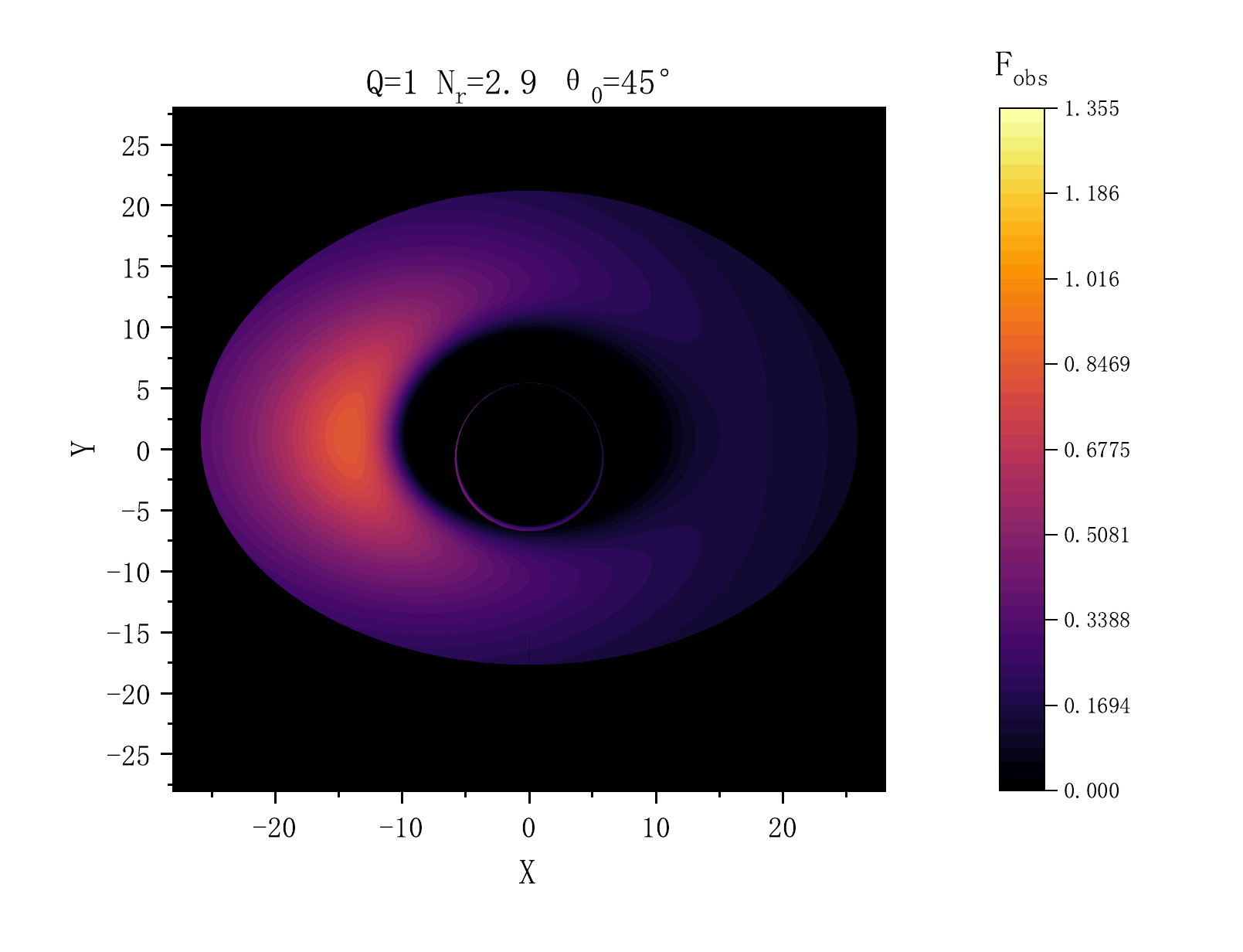}
\includegraphics[width=5cm,height=4.2cm]{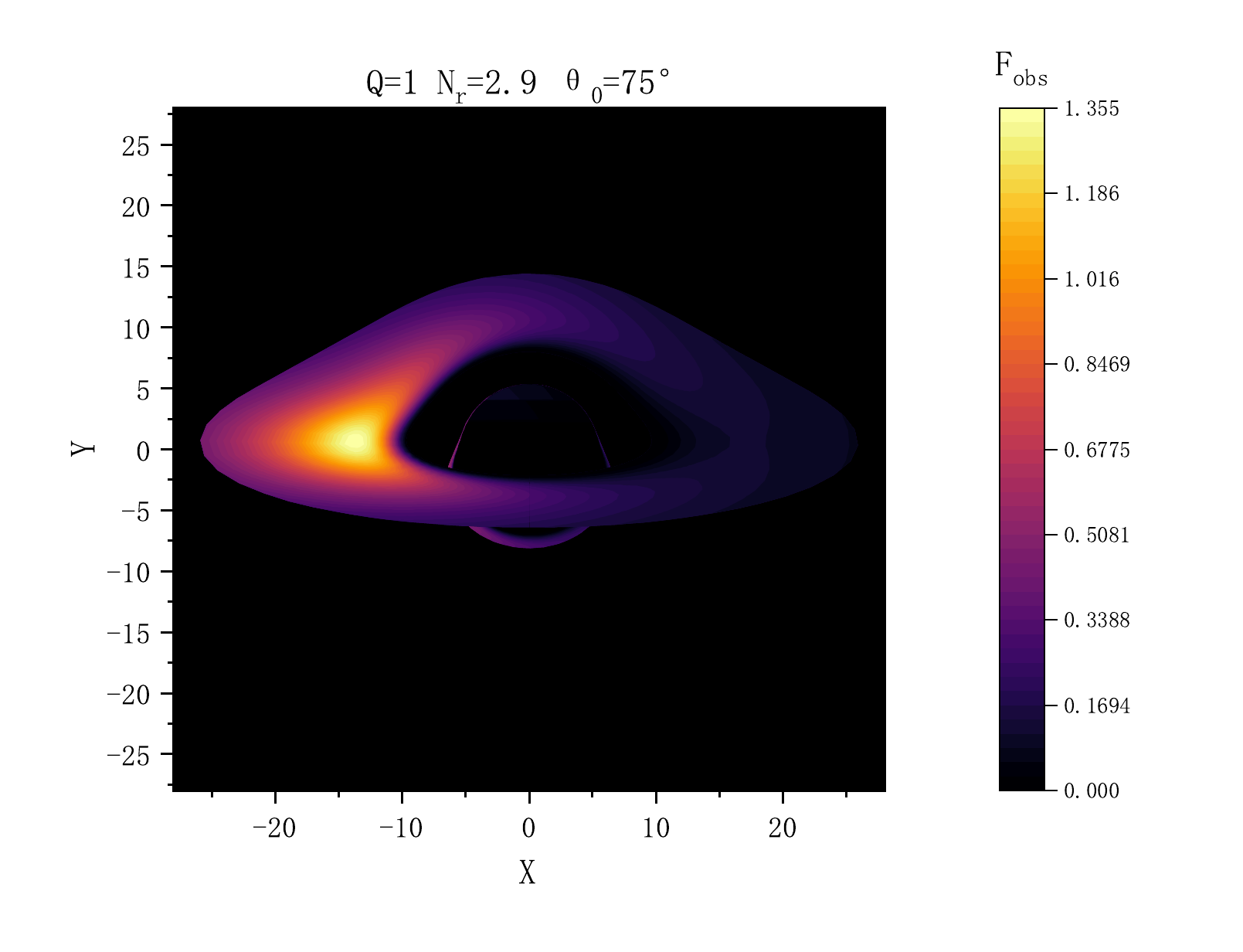}
\parbox[c]{15.0cm}{\footnotesize{\bf Fig~3.}  
The observed flux $F$ of the accretion disk is presented for three different observer tilt angles: $\theta_{0}=15^{\circ}$, $45^{\circ}$, and $75^{\circ}$. The black hole mass is $M = 1$, the charge is $Q=1$, and the radiation field parameter is $N_{\rm r}=2.9$. The inner edge of the disk at $r_{\rm in} = r_{\rm isco}$, and the outer edge of the disk is at $r=25M$.}
\label{fig3}
\end{center}

\subsection{Data analysis}
\label{sec:2-3}
\par
In this subsection, we delve into the behavior of a black hole enveloped by a perfect fluid radiation field, with a specific focus on two pivotal features: the photon ring ($r_{\rm ph}$) and the innermost stable circular orbit ($r_{\rm isco}$) of the black hole within the milieu of the perfect fluid radiation field. Figure 2 vividly demonstrates the discernible fluctuations in $r_{\rm isco}$ and $r_{\rm ph}$ of the accretion disk as we manipulate the parameters $Q$ and $N_{\rm r}$. To enhance the lucidity and furnish an intuitive exposition of our findings, we hold one parameter constant, either $Q$ ($Q=1$) or $N_{\rm r}$ ($N_{\rm r}=1$), while systematically varying the other parameter to acquire precise values for $r_{\rm isco}$ ($Ar_{\rm isco}$) and $r_{\rm ph}$ ($Ar_{\rm isco}$). This method allows us to isolate the impact of each parameter, facilitating a detailed examination of how they alter the intrinsic properties of the black hole, leading to changes in the spacetime structure. This, in turn, affects the geodesic paths of photon motion, ultimately resulting in differences in the positions of the photon ring and the innermost stable circular orbit.

\par
To accurately capture the inherent behavior and facilitate a comprehensive comprehension, we employ a numerical fitting technique employing an exponential function to model the acquired data. This fitting methodology demonstrates a robust relationship between parameters and the circular geodesics of spacetime metrics. By ascertaining precise fitting parameters, we can elucidate the functional dependencies between $Q$, $N_{\rm r}$, $r_{\rm isco}$, and $r_{\rm ph}$. This enables us to obtain valuable information regarding the interconnections among the black hole, perfect fluid radiation field, and the position of the accretion disk.

\par
The utilization of numerical fitting and the application of an exponential function in our analysis ensure precise and dependable representation of the data. The fitted curves facilitate interpolation and extrapolation of the behavior of $r_{\rm isco}$ and $r_{\rm ph}$ for various values of $Q$ and $N_{\rm r}$, thereby offering a comprehensive grasp of the system's behavior. This methodology enhances the comprehensibility of our research outcomes and aids in identifying trends and patterns in the variations of the accretion disk's position.

\begin{center}
\includegraphics[width=5cm,height=5cm]{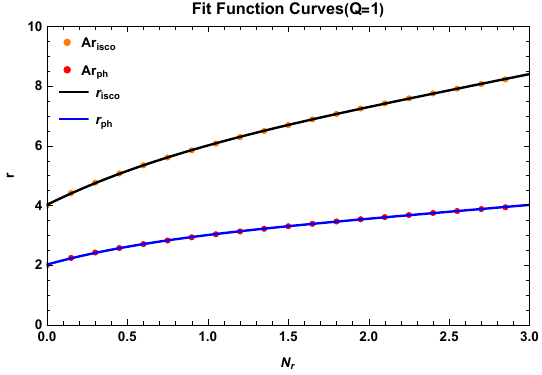}
\includegraphics[width=5cm,height=5cm]{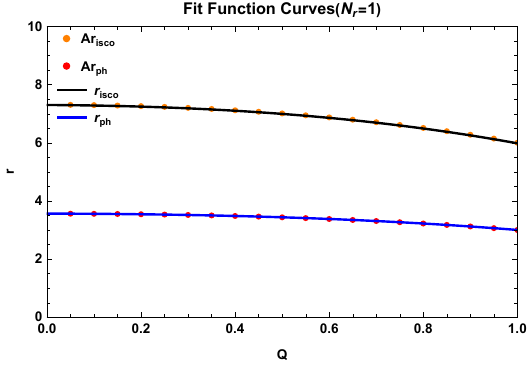}
\parbox[c]{15.0cm}{\footnotesize{\bf Fig~4.}
Exponential function fitting is applied to the black hole innermost stable orbit ($r_{\rm isco}$) and photon ring ($r_{\rm ph}$). In the {\em Left Panel}, the charge $Q$ is set to $1$, and the radiation field parameters $N_{\rm r}$ vary from $0$ to $2.99$. In the {\em Right Panel}, the radiation field parameters $N_{\rm r}$ is fixed at 1, and the charge $Q$ varies from $0$ to $1$.}
\label{fig4}
\end{center}

\par
In Figure 4, we observe the behavior of $r_{\rm isco}$ and $r_{\rm ph}$ while varying the parameters $Q$ and $N_{\rm r}$. When $Q$ is held constant, both $r_{\rm isco}$ and $r_{\rm ph}$ exhibit a continuous increase as $N_{\rm r}$ increases. Conversely, when $N_{\rm r}$ is fixed, an increase in $Q$ results in a continuous decrease in $r_{\rm isco}$ and $r_{\rm ph}$. Notably, it is discernible that changes in $N_{\rm r}$ have a more pronounced impact on $r_{\rm isco}$ and $r_{\rm ph}$. The figure presents two distinct curves, depicted in black and blue, along with their respective exponential fitting functions, described as follows:
\begin{equation}
\label{2-14}
F(x)= A E^{B x}+C E^{D x}.
\end{equation}
where $F(x)$ denotes either $r_{isco}$ or $r_{ph}$, and $x$ represents the varying parameter $N_{r}$ or $Q$. The values of each parameter in the fitting function are shown in Table 1.

\begin{table}[ht]
\begin{center}
\setlength{\tabcolsep}{1mm}
\linespread{0.1cm}
\begin{tabular}[t]{|c|c|c|c|c|}
  \hline
  $Function$ &${A}$ &${B}$ &${C}$ &${D}$ \\
  \hline
  $r_{\rm isco}(Q=1)$ &$5.977$       &$0.1173$       &$-1.957$     &$-1.013$   \\
  \hline
  $r_{\rm ph}(Q=1)$ &$2.898$       &$0.1103$       &$-0.8815$     &$-1.356$   \\
  \hline
  $r_{\rm isco}(N_{r}=1)$ &$-379.5$       &$0.4853$       &$386.8$     &$0.4759$   \\
  \hline
  $r_{\rm ph}(N_{r}=1)$ &$-1.502$       &$0.8227$       &$5.066$     &$0.237$   \\
  \hline
\end{tabular}
\end{center}
\caption{Exponential fitting function coefficients under various conditions.}\label{Table.1}
\end{table}

\section{Rastall parameters under strong gravitational lensing of galaxies}
\label{sec:3}
To rigorously constrain the gravitational parameters inherent to Rastall's theory within the framework of a perfect fluid, Li $et~al$ adopt a power-law mass density profile for Early Type Galaxies (ETGs) while incorporating Rastall gravity \cite{ccc}. This inquiry assumes spherical symmetry in mass distribution and characterizes the matter composition as a perfect fluid. In the context of galaxies, the mass density can be reasonably approximated as that of a perfect fluid, given that their dispersion velocity is markedly lower than the speed of light. In this scenario, the following equation may be utillized:
\begin{equation}
\label{3-1}
M(r_{\rm g})= 4 \pi \int r_{\rm g}^{2} \rho {\rm d}r,
\end{equation}
where $r_{\rm g}$ is the radius of the galaxy. By assuming a power law total mass density profile for a lens galaxy, the total mass enclosed within a sphere of radius $r_{\rm g}$ can be determined:
\begin{equation}
\label{3-2}
M(r_{\rm g})= \frac{2 \Gamma(\frac{1}{2}\gamma)}{\Gamma(\frac{1}{2}\gamma-\frac{1}{2}) \Gamma(\frac{1}{2})} D_{l}\theta_{\rm E}^{2}\frac{c^{2}}{4 G} \frac{D_{\rm s}}{D_{\rm ls}}R_{\rm ein}^{\gamma-3}r_{\rm g}^{3-\gamma},
\end{equation}
where $\theta_{\rm E}$ is the Einstein radius, $D_{\rm l}$ and $D_{\rm s}$ represent the angular-diameter distances of the lens and the source, respectively, $D_{\rm ls}$ denotes the angular-diameter distance between the lens and the source, $\gamma$ is the effective slope given by Treu and Koopmans \cite{ddd}. Due to the very similar power-law forms of Eq. (\ref{3-1}) and Eq. (\ref{3-2}), combining the two equations yields:
\begin{equation}
\label{3-3}
3-\gamma=\frac{3\kappa \lambda(1+\omega)-3\omega}{1-3\kappa \lambda(1+\omega)},
\end{equation}
By adjusting the parameter $\omega$, different formulations for $\kappa \lambda$ can be derived in various scenarios. Within the framework of Rastall gravity, the Weak Energy Conditions (WEC) ensure that the total energy density of any material field, as measured by an observer following a timelike trajectory, remains non-negative. Additionally, the convergence of geodesics imposes constraints consistent with the Strong Energy Conditions (SEC). Li $et~al.$ constrained the $\kappa \lambda$ Gaussian distribution function by employing $118$ gravitational lensing samples while setting $\omega$ to $0$ \cite{ccc}. Observational constraints were applied to $\kappa \lambda$ through both SEC and WEC, as summarized in Table 2, where $\alpha$ serves as an integration constant representing the local field structures.
\begin{table*}
\begin{center}
\setlength{\tabcolsep}{1mm}
\linespread{0.1cm}
\begin{tabular}[t]{|c|c|c|c|c|}
  \hline
  $Integration Constant$    &$Coupling Constant$    &$Observation Limitations$  \\
  \hline
  $\alpha > 0$         &$\kappa > 0$       &$ 0 \leq \kappa \lambda \leq \frac{1}{6}$          \\
  \hline
  $\alpha < 0 $         &$\kappa < 0 $       &$ \frac{1}{6} \leq \kappa \lambda \leq \frac{1}{4}$         \\
  \hline
\end{tabular}
\end{center}
\caption{Rastall gravity parameter limitations under early galaxies.}\label{Table.2}
\end{table*}

\par
In the specific scenario under investigation, involving a perfect fluid radiation field, there are noteworthy characteristics. When $\omega=\frac{1}{3}$, $\kappa \lambda$ remains invariant. At the scale of black holes, Rastall's gravitational theory could conceivably exert an influence on the mass and morphology of these cosmic entities. This influence may also yield novel predictions pertaining to black hole event horizons and the gravitational lensing phenomena associated with them. The observation of a black hole's gravitational lensing effect provides an avenue for constraining the Rastall parameter on the scale of black holes. It is essential to emphasize that a black hole enshrouded within a perfect fluid radiation field retains its exceptional nature within this framework.

\section{Conclusions and Discussion}
\label{sec:4}
\par
In summary, our invesgation provides a thorough analysis of the observable characteristics manifested by accretion disks surrounding black holes within the framework of Rastall gravity. In our study set against the backdrop of Rastall gravity, by comparing with Einstein's theory of gravity, we have examined a black hole encompassed by a perfect fluid radiation field. We have identified the influence of the observable properties of accretion disks stemming from the changes in the spacetime structure due to the variation of parameters within the Rastall gravity framework. This study has unveiled profound connections between the parameters $Q$ and $N_{\rm r}$ and the behavior of the accretion disk.

\par
First and foremost, we have established that the expansion or contraction of the accretion disk's orbit is intricately tied to the values of $Q$ and $N_{\rm r}$. Increasing the charge parameter $Q$ leads to an intensified gravitational pull from the black hole, thereby inducing a contraction of the accretion disk toward the black hole. Conversely, an increase in the parameter $N_{\rm r}$ triggers an outward expansion of the accretion disk due to the influence of the perfect fluid radiation field. Importantly, the visual appearance of a charged black hole within a perfect fluid radiation field in the Rastall theory bears no significant disparities when compared to the RN black hole. This observation suggests that, in terms of appearance, the effects of Rastall gravity and Einstein gravity are akin in this specific scenario.

\par
In addition, the brightness and stability characteristics of the accretion disk are intricately linked to the values of the parameters $Q$ and $N_{\rm r}$. An increase in the value of $N_{\rm r}$ is associated with a significant reduction in the brightness of the accretion disk. Furthermore, with the increase in $N_{\rm r}$, the position of the innermost stable circular orbit gradually expands outward. Conversely, keeping $N_{\rm r}$ constant while varying the parameter $Q$ results in a contrary trend in the position of the innermost stable circular orbit and the brightness of the accretion disk. These findings underscore the pivotal role of the perfect fluid radiation field parameter $N_{\rm r}$ and charge $Q$ in shaping the luminosity and stability properties of the accretion disk. Moreover, our analysis of the observer's perspective unveils morphological transformations in the accretion disk as the observation angle increases. The disk undergoes a transformation from a disk-like configuration to a hat-like structure, accompanied by a growing degree of asymmetry in the radiant flux intensity on the left and right sides of the disk. This asymmetry can be attributed to the heightened influence of the Doppler effect, leading to a more pronounced red or blue shift in the observed radiation as the observation angle rises.

\par
Astronomical observations wield a critical role in the rigorous constraint of the gravitational parameters intrinsic to Rastall's theory. By scrutinizing various astrophysical phenomena, such as gravitational lensing effects, black hole properties, and other observational data, we gain valuable insights and constraints on the specific gravitational parameters associated with Rastall's theory. These observations serve as a cornerstone for refining our comprehension of gravity and its behavior on astronomical scales.

\par
In conclusion, our investigation furnishes valuable insights into the observable attributes characterizing accretion disks encircling black holes within the framework of Rastall gravity. This underscores the close correlation between the parameters $Q$ and $N_{\rm r}$ and the behavior of the accretion disk, revealing unique signatures and implications of Rastall gravity in observational properties of these celestial systems. These findings contribute substantially to a deeper understanding of the intricate interplay among black holes, perfect fluid radiation fields, and the resulting properties of accretion disks.

\section*{Acknowledgments}
This work is supported by the National Natural Science Foundation of China (Grant No. 11805166) and the Sichuan Youth Science and Technology Innovation Research Team (Grant No. 21CXTD0038).

\clearpage

\end{CJK}

\section{References}
\addcontentsline{toc}{chapter}{References}
\begin{thebibliography}{99}\footnotesize
\itemsep=-3pt plus.2pt minus.2pt   

\bibitem{1}
P. Rastall. Generalization of the einstein theory. Phys. Rev. D 1972; \textbf{6}: 3357-3359.

\bibitem{2}
A. M. Oliveira, H.E.S. Velten,J.C. Fabris, et al. Neutron Stars in Rastall Gravity.Phys.Rev.D 2015; \textbf{92}: 044020.

\bibitem{3}
Y. Heydarzade, F. Darabi. Black Hole Solutions Surrounded by Perfect Fluid in Rastall Theory. Phys.Lett.B 2017; \textbf{771}: 365-373.

\bibitem{4}
H. Moradpour, N. Sadeghnezhad, S.H. Hendi.Traversable asymptotically flat wormholes in Rastall gravity. Can.J.Phys. (2017); \textbf{95}:  1257-1266.

\bibitem{5}
R. Kumar, S. Ghosh. Rotating black hole in Rastall theory. Eur.Phys.J.C 2018; \textbf{78}: 750.

\bibitem{6}
M. Visser. Rastall gravity is equivalent to Einstein gravity. Phys.Lett.B 2018; \textbf{782}: 83-86.

\bibitem{7}
F. Darabi, H. Moradpour, I. Licata, et al. Einstein and Rastall Theories of Gravitation in Comparison. Eur.Phys.J.C 2018; \textbf{78}: 25.

\bibitem{8}
G. Abbas, M. R. Shahzad. A new model of quintessence compact stars in the Rastall theory of gravity. Eur.Phys.J.A 2018; \textbf{54}: 211.

\bibitem{9}
D. Das, S. Dutta, S. Chakraborty. Cosmological consequences in the framework of generalized Rastall theory of gravity. Eur.Phys.J.C 2018; \textbf{78}: 810.

\bibitem{10}
S. Hansraj, A. Banerjee, P. Channuie. Impact of the Rastall parameter on perfect fluid spheres. Annals Phys. 2019; \textbf{400}: 320-345.

\bibitem{11}
X. C. Cai, Y. G. Miao. Quasinormal modes and spectroscopy of a Schwarzschild black hole surrounded by a cloud of strings in Rastall gravity. Phys.Rev.D 2020; \textbf{101}: 104023.

\bibitem{12}
Z. Li, T. Zhou. Kerr black hole surrounded by a cloud of strings and its weak gravitational lensing in Rastall gravity. Phys.Rev.D 2021; \textbf{104}: 104044.

\bibitem{13}
K. Akiyama et al. [Event Horizon Telescope Collaboration], First M87 Event Horizon Telescope Results. I. The Shadow of the Supermassive Black Hole. Astrophys. J.
L 2019; \textbf{1}: 875.

\bibitem{14}
K. Akiyama et al, [Event Horizon Telescope Collaboration], First Sagittarius A* Event Horizon Telescope Results. I. The Shadow of the Supermassive Black Hole
in the Center of the Milky Way. Astrophys. J. Lett. L 2022; \textbf{12}: 930.

\bibitem{15}
N. I. Shakura, R. A. Sunyaev, Black holes in binary systems. Observational appearance, Astron. Astrophys 1973; \textbf{24}: 337-355.

\bibitem{16}
J. -P. Luminet, Image of a spherical black hole with thin accretion disk, Astron. Astrophys. 1979; \textbf{75}: 1.

\bibitem{17}
A. Laor. Line profiles from a disk around a rotating black hole. Astrophys. J 1991; \textbf{376}: 90.

\bibitem{18}
G. Gyulchev, P. Nedkova, T. Vetsov, S. Yazadjiev, Image of the Janis-Newman-Winicour naked singularity with a thin accretion disk, Phys. Rev. D 2019; \textbf{100}: 024055.

\bibitem{19}
S. Paul, R. Shaikh, P. Banerjee, T. Sarkar, Observational signatures of wormholes with thin accretion disks, JCAP 2020; \textbf{03}: 055.

\bibitem{20}
F. Rahaman, T. Manna, R. Shaikh, S. Aktar, M. Mondal, Thin accretion disks around traversable wormholes, Nucl. Phys. B 2021; \textbf{972}: 115548.

\bibitem{21}
C. Q. Liu, L. Tang, J. L. Jing, Image of the Schwarzschild black hole pierced by a cosmic string with a thin accretion disk, Int. J. Mod. Phys. D 2022; \textbf{31}: 2250041.

\bibitem{nnn}
K. Meng, X. L. Fan, S. L, Dynamics of null particles and shadow for general rotating black hole, arXiv: 2307.08953.

\bibitem{22}
S. Guo, K. J. He ,G. R. L, et al. The shadow and photon sphere of the charged black hole in Rastall gravity. Class.Quant.Grav. 2021; \textbf{38}: 165013.

\bibitem{23}
S. Guo, G. R. Li, E. W. Liang. Observable characteristics of the charged black hole surrounded by thin disk accretion in Rastall gravity. Class. Quant. Grav. 2022; \textbf{39}: 135004.

\bibitem{eee}
F. Rahaman, Peter K. F. Kuhfittig, K. Chakraborty, et al. Modeling galactic halos with predominantly quintessential matter, Int. J. Theor. Phys. 2011; \textbf{50}: 2655-2665.

\bibitem{fff}
F. Rahaman, K. K. Nandi, A. Bhadra, et al. Perfect Fluid Dark Matter, Phys. Lett. B. 2011; \textbf{694}: 10-15.

\bibitem{aaa}
V.V. Kiselev. Quintessence and black holes. Class. Quant. Grav. 2003; \textbf{20}: 1187-1198

\bibitem{bbb}
A. Vikman. Can dark energy evolve to the phantom? Phys. Rev. D. 2005; \textbf{71}: 023515

\bibitem{24}
I. D. Novikov, K. S. Thorne, in ¡°Black Holes¡±, ed. C. DeWitt and B. DeWitt, New York: Gordon and Breach 1973.

\bibitem{25}
D. N. Page, K. S. Thorne, Disk-Accretion onto a Black Hole. Time-Averaged Structure of Accretion Disk, Astrophys. J. 1974; \textbf{191}: 499.

\bibitem{ccc}
R. Li, J. Wang, Z. Xu, et al. Constraining the Rastall parameters in static space¨Ctimes with galaxy-scale strong gravitational lensing, Mon. Not. Roy. Astron. Soc. 2019; \textbf{486}: 2407-2411.

\bibitem{ddd}
L.V.E. Koopmans, T. Treu, The structure and dynamics of luminous and dark matter in the early-type lens galaxy of 0047-281 at z=0.485, Astrophys. J. 2003; \textbf{583}: 606-615.



\end{thebibliography}
\end{document}